\documentclass[prb,twocolumn,floatfix,showpacs,showkeywords]{revtex4}
\usepackage{amssymb}
\usepackage{amsmath}
\usepackage{amsfonts}
\usepackage{bm}
\usepackage{graphicx}

\newcommand{\llangle}{\langle\kern-.25em\langle}
\newcommand{\rrangle}{\rangle\kern-.25em\rangle}

\begin{document}

\title{Phase diagram of the frustrated, spatially anisotropic $S=1$
antiferromagnet on a square lattice}
\author{H.~C. Jiang$^1$}
\author{F. Kr\"uger$^2$}
\author{J. E. Moore$^3$}
\author{D.~N. Sheng$^4$}
\author{J. Zaanen$^5$}
\author{Z.~Y. Weng$^1$}
\affiliation{$^1$Center for Advanced Study, Tsinghua University, Beijing, 100084, China\\
$^2$Department of Physics, University of Illinois, 1110 W. Green St.,
Urbana, IL 61801, USA\\
$^3$Department of Physics, University of California, Berkeley, CA 94720, USA%
\\
$^4$Department of Physics and Astronomy, California State University,
Northridge, California 91330, USA\\
$^5$Instituut-Lorentz, Universiteit Leiden, P. O. Box 9506, 2300 RA Leiden,
The Netherlands}
\date{\today}

\begin{abstract}
We study the $S=1$ square lattice Heisenberg antiferromagnet with spatially
anisotropic nearest neighbor couplings $J_{1x}$, $J_{1y}$ frustrated by a
next-nearest neighbor coupling $J_{2}$ numerically using the density-matrix
renormalization group (DMRG) method and analytically employing the
Schwinger-Boson mean-field theory (SBMFT). Up to relatively strong values of
the anisotropy, within both methods we find quantum fluctuations to
stabilize the N\'{e}el ordered state above the classically stable region.
Whereas SBMFT suggests a fluctuation-induced first order transition between
the N\'{e}el state and a stripe antiferromagnet for $1/3\leq
J_{1x}/J_{1y}\leq 1$ and an intermediate paramagnetic region opening only
for very strong anisotropy, the DMRG results clearly demonstrate that the
two magnetically ordered phases are separated by a quantum disordered region
for all values of the anisotropy with the remarkable implication that the
quantum paramagnetic phase of the spatially isotropic $J_{1}$-$J_{2}$ model
is continuously connected to the limit of decoupled Haldane spin chains. Our
findings indicate that for $S=1$ quantum fluctuations in strongly frustrated
antiferromagnets are crucial and not correctly treated on the semiclassical
level.
\end{abstract}

\maketitle

\section{Introduction}

A striking commonality between the recently discovered iron pnictides
superconductors\cite{Kamihara+08,Takahashi+08,Ren+08,Chen+08,Chen+08b}
and the high-$T_c$ cuprates is that in both cases
superconductivity emerges on doping antiferromagnetic parent compounds. In
trying to unravel the mechanisms of superconductivity frustrated quantum
antiferromagnets on the square lattice have been subject to intense research
over the last decades with particular interest in the extreme quantum limit $%
S=\frac 12$ relevant for the cuprates, whereas the interest in higher spin
values increased tremendously with the discovery of the pnictides.

Thereby, the Heisenberg model with antiferromagnetic exchange couplings $%
J_{1}$ and $J_{2}$ between nearest neighbor (NN) and next-nearest neighbor
(NNN) has served as a prototype model for studying magnetic frustration. On
a classical level, one finds N\'{e}el order to be stable for $%
J_{2}/J_{1}\leq 1/2$ whereas for larger ratios the classical ground state is
give by a stripe-antiferromagnet with ordering wave-vector $(\pi ,0)$. Not
surprisingly, the $J_{1}$-$J_{2}$ model has been used to rationalize the $%
(\pi ,0)$ magnetism\cite{Cruz+08} of the iron pnictide superconductors and to
subsequently calculate the magnetic excitation spectra\cite{Yao+08,Zhao+08}
where the incorporation of a strong anisotropy between the NN couplings turned out to
be necessary to reproduce the low energy spin-wave excitations. Recently, it has been
suggested that the strong anisotropies in the magnetism\cite{Zhao+08}, but also
in electronic properties\cite{McGuire+08} of the pnictides originate in the coupling
to orbital degrees of freedom arising from an orbital degeneracy of an intermediate
$S=1$ spin state\cite{Kruger+08}. Even if the charge degrees of freedom are not completely
localized, the $J_{1}$-$J_{2}$ model can be used as the starting point for a symmetry-based
analysis of magnetism and superconductivity in the iron-based superconductors~\cite%
{Xu+08,Hu+08}.

The $1/S$ expansion serves as a natural starting point to investigate the
stability of the classical orders against quantum fluctuations which are
expected to induce a paramagnetic phase near $J_2/J_1= 1/2$ where both
orders compete. Indeed, on the level of lowest order linear spin-wave theory
one finds an intermediate paramagnetic phase for all spin values\cite%
{Chandra+88}. However, the incorporation of spin-wave interactions within a
modified spin-wave theory (MSWT)\cite{Xu+90,Gochev95} or the SBMFT\cite%
{Mila+91} drastically changes this picture. Within both approaches one finds
a dramatic stabilization of the classical orders by quantum fluctuation for
the N\'eel order even up to values considerably larger than the classical
threshold value $J_2/J_1= 1/2$. This has been interpreted as an
order-out-of-disorder phenomenon giving rise to a fluctuation induced first
order transition between the two orders. This picture has been confirmed
recently by a functional one-loop renormalization group analysis\cite%
{Kruger+06} where it was shown that the N\'eel phase becomes unstable
towards a fluctuation induced first order transition for $S>0.68$.

Surely, it is questionable if the semiclassical treatment can correctly
account for the quantum fluctuations in the regime of strong frustration and
small spins. In fact, in the case $S=\frac 12$ various numerical studies
including exact diagonalization\cite%
{Dagotto+89,Schulz+92,Schulz+96,Sindzingre04}, variational Monte Carlo\cite%
{Capriotti+00,Capriotti+01} series expansion\cite%
{Gerlfand+89,Gerlfand90,Singh+99,Sushkov+01,Sushkov+02} as well as the
coupled cluster approach\cite{Bishop+98} give the consistent picture that in
the regime $0.4\lesssim J_2/J_1\lesssim 0.6$ no magnetic order is present
clearly indicating that the aforementioned semiclassical treatments
overestimate the stability of the ordered states. Another key observation of
the series expansion and in particular of the unbiased exact diagonalization
studies is that in the paramagnetic phase the lattice symmetry is
spontaneously broken due to the formation of columnar valence bond solid
order. Such states have been predicted before\cite{Read+89,Read+90} on the
basis of a large $N$ treatment for non frustrated systems in the absence of
long-range antiferromagnetic order. These depend however crucially on the value of the
spin, revealing the subtle workings of the spin Berry phases (see e.g. [\onlinecite{Sachdev08}]
and references therein). For $S=\frac 12$  these act to give a special stability to
'valence bond'  pair singlets involving nearest-neighbor spins,
and the quantum disordered phases are actually valence bond crystals breaking the translational
symmetry of the square lattice further to fourfold degenerate spin-Peierls states.  But for $%
S=1$ the Berry phases act in favor of the formation of 'chain singlets'\cite{Read+89,Read+90}
that can stack either along the $x$ or $y$ direction of the square lattice yielding a twofold
degenerate ground state.

\begin{figure}[h]
\includegraphics[width=0.95\linewidth]{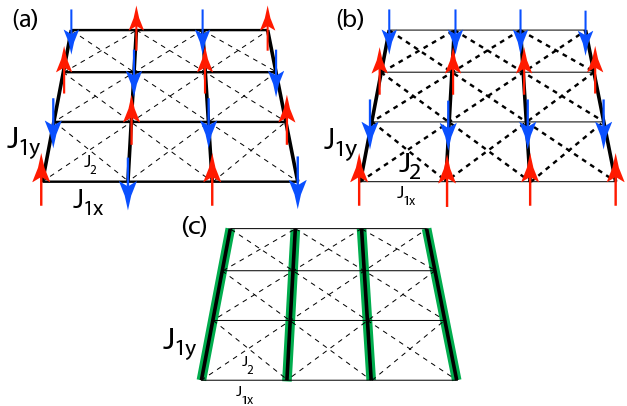}
\caption{The model studied in this paper consists of  a $S=1$ Heisenberg system with
spatially anisotropic nearest-neighbor couplings $J_{1x}$, $J_{1y}$ and isotropic
next-nearest-neighbor couplings $J_2$. Pending the balance of these couplings one finds either
a simple staggered- (a) or 'stripe' (b) antiferromagnetic order.  In the limit of isolated chains
a stack of isolated Haldane spin chains is formed (c) and based on our DMRG calculations we
could conclude that these survive all the way to the isotropic limit in the vicinity of the point of
maximal frustration ($J_2/J_1 =0.5$).}
\label{Fig:intro}
\end{figure}

In this paper we investigate the $S=1$ version of the spatially anisotropic $%
J_1$-$J_2$ model, given by the Hamiltonian
\begin{eqnarray}
H=J_{1x}\sum_{\langle i,j \rangle_x}\mathbf{S}_i \mathbf{S}_j+
J_{1y}\sum_{\langle i,j \rangle_y}\mathbf{S}_i \mathbf{S}_j+
J_{2}\sum_{\langle\langle i,j \rangle\rangle}\mathbf{S}_i \mathbf{S}_j,
\label{Hamiltonian}
\end{eqnarray}
where the first two sums run over NN spins with exchange couplings $%
J_{1x}\le J_{1y}$ in $x$ and $y$ directions, respectively, and the third sum
runs over all NNN pairs with exchange couplings $J_2$ as illustrated in Fig~%
\ref{Fig:intro}. We note that both the spatially isotropic $J_1$-$J_2$ model
as well as the decoupled Haldane spin-chain limit appear as special cases of
Hamiltonian (\ref{Hamiltonian}).

Unlike for $S=\frac 12$ this model has been hardly explored for $S=1$
despite the potential relevance for the iron pnictides, and to best of our
knowledge exact diagonalization studies are not available. In a very recent
coupled cluster treatment\cite{Bishop+08} it has been found that at the
isotropic point $J_{1x}=J_{1y}=J_1$ the N\'eel and stripe ordered phases are
separated by a first order transition at $J_2/J_1\approx 0.55$ slightly
smaller than the MSWT\cite{Xu+90,Gochev95} and SBMFT\cite{Mila+91} results
and that the transition remains first order up to an anisotropy $%
J_{1x}/J_{1y}\approx 0.66$. For stronger anisotropies the authors find
continuous transitions very close to the classical transition line $%
J_2=J_{1x}/2$, although they were not able to resolve an intermediate
paramagnetic region within the numerical resolution. On contrary, a two step
DMRG study\cite{Moukouri06} at the particular point $J_{1x}/J_{1y}=0.2$
indicated a much wider non-magnetic region with a spin-gap reaching its
maximum $\Delta\approx 0.39 J_{1y}$ close to the maximally frustrated point $%
J_2/J_{1y}=0.1$, which is only slightly below the gap $\Delta_H\approx 0.41
J $ [\onlinecite{Sorenson+93}] of an isolated Haldane chain\cite%
{Haldane83a,Haldane83b}.

Starting from the Haldane chain limit ($J_{1x}=J_{2}=0$) it has been
predicted by Monte Carlo simulations\cite{Todo+01} as well as analytically%
\cite{Anfuso+07} that an infinitesimal coupling $J_{1x}$ leads to the
destruction of the topological string order\cite{Nijs+89} of the Haldane chain. 
However, due to the protection by the finite energy gap of the
spin-1 Haldane chain all other ground-state properties as well as the
thermodynamics are only minimally affected by a small interchain coupling $%
J_{1x}$ and a finite, albeit small\cite{Kim+00,Matsumoto+01,Alet+02,Zinke+08}
coupling is necessary to establish magnetic long range order. Whereas a
similar reasoning should hold for a vanishing NN coupling perpendicular to
the chains ($J_{1x}=0$) and small diagonal coupling $J_{2}$ the Haldane
chain phase seems to be considerably more stable against the simultaneous
increase of the two mutually frustrating couplings $J_{1x}$ and $J_{2}$ as
indicated by two step DMRG results showing an almost negligible reduction of
the spin gap for moderate interchain couplings\cite{Moukouri06}. This
suggests that the paramagnetic phase of the frustrated two dimensional spin
model can be viewed as a continuation of the Haldane spin chain. Recently,
based on a theoretical method with a continuous deformation of the $J_{1}$-$%
J_{2}$ model, the ground state at the special isotropic case
$J_{2}=J_{1}/2$ has been conjectured\cite{cai+07} to be a two-fold
degenerate valence bond solid state along either the horizontal or
vertical direction of the square lattice. The main result of
this paper is that by employing DMRG and by studying systematically
how matters evolve as function of increasing anisotropy we arrive at
solid evidence for the smooth continuation between the Haldane chain
phase and an isotropic disordered phase near $J_{2}=J_{1}/2$.

Since for $S=1$ exact diagonalization is restricted to very small system
sizes and since the quantum Monte Carlo method suffers from the infamous
sign problem in the presence of frustration we employ the DMRG method\cite%
{White92} to map out the phase diagram of the spatially anisotropic $%
J_{1x}-J_{1y}-J_2$ model (\ref{Hamiltonian}) over the whole parameter range
including the decoupled Haldane spin chains and the isotropic $J_1$-$J_2$
model as limiting cases. The DMRG has the merits of not being biased as
analytical treatments based on $1/S$ or $1/N$ expansions, being capable of
dealing with frustrated systems, and allowing to investigate much bigger
system than possible with exact diagonalization. Moreover, it is capable of
reproducing the spin gap of the decoupled Haldane chain limit with high
accuracy\cite{Moukouri06}. The resulting phase diagram is contrasted by
analytical results we obtain within a generalization of the SBMFT to the
anisotropic system.

The remainder of the paper is organized as follows. In Sec.~\ref{sec.SBMFT}
we outline the SBMFT and calculate the resulting phase diagram which is used
for comparison with the numerical results. In Sec.~\ref{sec.DMRG} we use the
DMRG to construct the phase diagram by a careful finite-size analysis of the
magnetic structure factor at the ordering wave vectors of the two
magnetically ordered phases as well as of the spin-gap in the non magnetic
region as a function of the NNN coupling $J_{2}$ for various ratios of the
NN couplings $J_{1x}$ and $J_{1y}$. Finally, in Sec.~\ref{sec.discussion}
the results obtained within the two methods are compared and the
shortcomings of the semiclassical approach are highlighted. Further, the 
potential relevance of our findings for the iron pnictides is discussed.

\section{SBMFT}

\label{sec.SBMFT}

In this section we generalize the SBMF calculation for the isotropic $J_1$-$%
J_2$-model\cite{Mila+91} to the case of anisotropic nearest-neigbor exchange
couplings. For abbreviation we define $J_x:=J_{1x}$, $J_y:=J_{1y}$, and $%
J_d:=J_2$. Assuming $J_x\le J_y$, the classical ground states are given by a
N\'eel ordered state with an ordering wavevector ${\bm Q}%
=(\pi,\pi)$ for $J_d/J_x\le1/2$ and by a columnar antiferromagnetic state
with ${\bm Q}=(0,\pi)$ for $J_d/J_x>1/2$. Following the previous
calculations, we perform a spin rotation ${\tilde{S}}^x_i=\sigma_i S^x_i$, ${%
\tilde{S}}^y_i=\sigma_i S^y_i$, ${\tilde{S}}^z_i=S^z_i$, where $%
\sigma_i=\exp(i{\bm Q}{\bm r}_i)=\pm 1$ with ${\bm Q}$ the ordering
wavevectors of the two classical orders, and represent the rotated spin
operators in terms of Schwinger bosons $b_{i,\uparrow}$, $b_{i,\downarrow}$
as

\begin{equation}
{\tilde{\bm S}}_{i}=\frac{1}{2}b_{i,\nu }^{\dagger }{\boldsymbol{\sigma }}%
_{\nu ,\nu ^{\prime }}b_{i,\nu ^{\prime }}.
\end{equation}%
Here ${\boldsymbol{\sigma }}=(\sigma ^{x},\sigma ^{y},\sigma ^{z})$ with $%
\sigma ^{\alpha }$ the standard Pauli matrices. The constraint $b_{i,\nu
}^{\dagger }b_{i,\nu }=2S$ ensures that $\langle {\bm S}^{2}\rangle =S(S+1)$%
. The Hamiltonian (\ref{Hamiltonian}) can then be rewritten in the compact
form

\begin{eqnarray}
{\mathcal{H}} &=&\frac{1}{2}\sum_{(i,j)}J_{ij}\left[ \frac{\sigma _{i}\sigma
_{j}+1}{2}(F_{ij}^{\dagger }F_{ij}-2S^{2})\right.  \notag \\
&&\left. +\frac{\sigma _{i}\sigma _{j}-1}{2}(G_{ij}^{\dagger }G_{ij}-2S^{2})%
\right] ,  \label{HamSB}
\end{eqnarray}%
where the sum runs over all bonds and we have introduced the bond operators $%
F_{ij}^{\dagger }=b_{i,\nu }^{\dagger }b_{j,\nu }$ and $G_{ij}^{\dagger
}=b_{i,\nu }^{\dagger }b_{j,-\nu }^{\dagger }$. A mean-field decoupling is
then performed with respect to the order parameters $f_{ij}=\langle
F_{ij}^{\dagger }\rangle /2$ and $g_{ij}=\langle G_{ij}^{\dagger }\rangle /2$%
. For the N\'{e}el ordered phase we have to introduce fields $g_{x}$, $g_{y}$
for the non-frustrated nearest-neigbor bonds and $f_{d}$ for frustrated NNN
bonds whereas the $(0,\pi )$-phase is characterized by order-parameter
fields $g_{y}$ and $g_{d}$ for the non-frustrated bonds along the y and
diagonal directions and $f_{x}$ for the frustrated x-bonds. The local
constraints are replaced by a global one and treated with a Lagrange
multiplier $\lambda $. The resulting mean-field Hamiltonian, which in
momentum space is given by

\begin{eqnarray}
{\mathcal{H}}_{\text{mf}} & = & \int_{\bm q}\left[(h_{\bm %
q}+\lambda)(b^\dagger_{{\bm q} \uparrow}b_{{\bm q}\uparrow}+b^\dagger_{-{\bm %
q} \downarrow}b_{-{\bm q}\downarrow})\right.  \notag \\
& & \left.-\Delta_{\bm q}(b^\dagger_{{\bm q} \uparrow}b^\dagger_{-{\bm q}%
\downarrow}+b_{{\bm q} \uparrow}b_{-{\bm q}\downarrow})\right],
\label{HamMF}
\end{eqnarray}
is easily diagonalized by a Bogolioubov transformation yielding the
dispersion

\begin{equation}
\omega_{\bm q}=[(h_{\bm q}+\lambda)^2-\Delta_{\bm q}^2]^{1/2}
\end{equation}
with

\begin{subequations}
\begin{eqnarray}
h_{\bm q} & = & 4f_d J_d \cos q_x\cos q_y, \\
\Delta_{\bm q} & = & 2(g_x J_x\cos q_x+g_y J_y\cos q_y)
\end{eqnarray}
\end{subequations}
in the N\'eel phase and likewise in the $(0,\pi)$ phase

\begin{subequations}
\begin{eqnarray}
h_{\bm q} & = & 2 f_x J_x \cos q_x, \\
\Delta_{\bm q} & = & 2 g_y J_y\cos q_y+4 g_d J_d\cos q_x\cos q_y.
\end{eqnarray}
\end{subequations}

The Lagrange multiplier in the ordered phases is determined by the
requirement that $\omega_{\bm Q}=0$ for the corresponding ordering
wavevectors yielding $\lambda=2(g_x J_x+g_y J_y)-4f_d J_d$ in the N\'eel and
$\lambda=2(g_y J_y-f_x J_x)+4g_d J_d$ in the $(0,\pi)$-phase. The reduced
moment $S^*$ in the ordered phases is determined by

\begin{equation}
S^*=S-\frac 12 \left(\int_{\bm q} \frac{h_q+\lambda}{\omega_q}-1\right).
\end{equation}
Naturally, one might expect that quantum fluctuations tend to destabilize the
classical orders introducing an intermediate paramagnetic phase. In this
case the second order transitions between the two different magnetic orders
are determined by the lines where the corresponding magnetizations go to
zero, $S^*\to 0$. However, the analysis of the isotropic model\cite{Mila+91} 
($J_x=J_y$) shows that the quantum fluctuations can lead to a significant stabilization
of the N\'eel order leading to a region where $S^*_{(\pi,\pi)}>0$ and $%
S^*_{(0,\pi)}>0$. In this region which is expected to persist for not too
strong anisotropy, a discontinuous first order transition between the two
different magnetic orders is likely and can be estimated from a comparison
of the ground-state energies which immediately follow from Eq. (\ref{HamSB})
as

\begin{subequations}
\begin{eqnarray}
E_{(\pi,\pi)} & = & -J_x(2g_x^2-S^2)-J_y(2g_y^2-S^2)  \notag \\
& & +2J_d(2f_d^2-S^2) \\
E_{(0,\pi)} & = & J_x(2f_x^2-S^2)-J_y(2g_y^2-S^2)  \notag \\
& & -2J_d(2g_d^2-S^2).
\end{eqnarray}
\end{subequations}

The order parameter fields entering the mean-field Hamiltonian (\ref{HamMF})
have to be determined self consistently. Using the above Bogoliubov
transformation we obtain for the N\'eel ordered phase

\begin{subequations}
\begin{eqnarray}
g_{x} &=&S^{\ast }+\int_{\bm q}\frac{\Delta _{\bm q}}{2\omega _{\bm q}}\cos
q_{x}, \\
g_{y} &=&S^{\ast }+\int_{\bm q}\frac{\Delta _{\bm q}}{2\omega _{\bm q}}\cos
q_{y}, \\
f_{d} &=&S^{\ast }+\int_{\bm q}\frac{h_{\bm q}+\lambda }{2\omega _{\bm q}}%
\cos q_{x}\cos q_{y},
\end{eqnarray}
\end{subequations}
whereas in the $(0,\pi )$ phase the self-consistency equations for the
order-parameter fields are given by

\begin{subequations}
\begin{eqnarray}
f_x & = & S^*+\int_{\bm q}\frac{h_{\bm q}+\lambda}{2\omega_{\bm q}}\cos q_x,
\\
g_y & = & S^*+\int_{\bm q}\frac{\Delta_{\bm q}}{2\omega_{\bm q}}\cos q_y, \\
g_d & = & S^*+\int_{\bm q}\frac{\Delta_{\bm q}}{2\omega_{\bm q}}\cos q_x\cos
q_y.
\end{eqnarray}
\end{subequations}

\begin{figure}[h]
\includegraphics[width=0.95\linewidth]{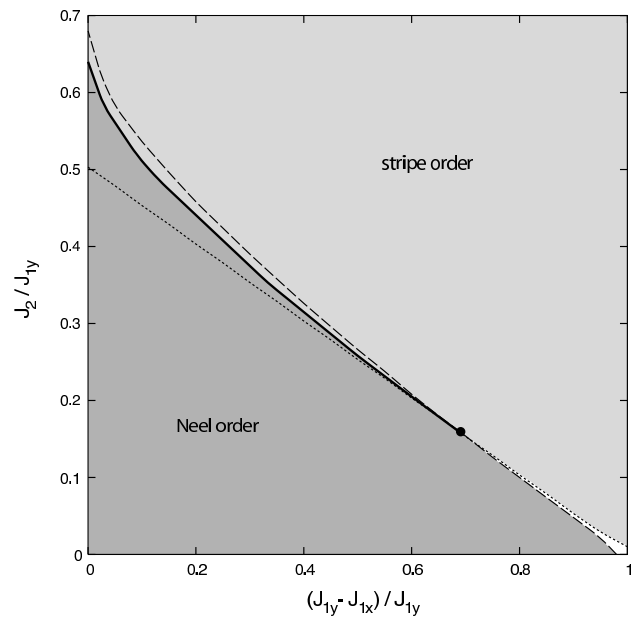}
\caption{Phase diagram for $S=1$ as a function of and anisotropy $\protect%
\alpha =(J_{1y}-J_{1x})/J_{1y}>0$ between the nearest neigbor exchange
couplings and relative strength of the next nearest neighbor coupling $%
J_{2}/J_{1y}$ obtained within SBMFT. Up to an anisotropy $\protect\alpha %
\approx 0.66$ the $(\protect\pi ,\protect\pi )$ and $(0,\protect\pi
)$ antiferromagnetic orders are separated by a first order
transition (solid line). At the tricritical point the first-order
line splits into two second-order lines (dashed and dotted)
separating the two magnetic phases from a gapped non-magnetic
phase.} \label{Fig:SBMFT}
\end{figure}

The resulting phase diagram for $J_{1x}\leq J_{1y}$ is shown in Fig. \ref%
{Fig:SBMFT} as a function of the anisotropy $\alpha =(J_{1y}-J_{1x})/J_{1y}$
between the NN exchange couplings and of the relative strength $\beta
=J_{2}/J_{1y}$ of the NNN coupling. In agreement with earlier studies of the
isotropic limit $J_{1x}=J_{1y}=J_{1}$ within SBMFT and MSWT we find a
dramatic stabilization of the N\'{e}el order above the classical value and a
large region $0.51\lesssim \beta \lesssim 0.68$ where the two competing
orders are potentially stable, indicated by $S_{(\pi ,\pi )}^{\ast }>0$ and $%
S_{(0,\pi )}^{\ast }>0$. The crossing of the self-consistently determined
energies of the two states suggests a first order transition at $\beta
\approx 0.64$ considerably larger than the classical value $1/2$ and about
10 percent bigger than the coupled cluster result\cite{Bishop+08}. Although
the region of coexistence is considerably narrowed by a small anisotropy we
find it to persist up to $\alpha \approx 0.66$ where the first order line
terminates. The existence of such a tricritical point was also suggested by
the coupled cluster analysis although located at a much smaller anisotropy $%
\alpha \approx 0.34$. For anisotropies $\alpha >0.66$ we find an
intermediate non-magnetic region ($S_{(\pi ,\pi )}^{\ast }=S_{(0,\pi
)}^{\ast }=0$) separating the two ordered phases. This region is found to be
very narrow and close to the classical phase boundary $\beta =(1-\alpha )/2$.

Although the existence of a tricritical point and of a very narrow
paramagnetic strip terminating at the Haldane chain limit $J_{1x}=J_2=0$ is
in qualitative agreement with the coupled cluster results\cite{Bishop+08}
the small width of the paramagnetic region is in disagreement with other
available numerical results. For $J_{1x}/J_{1y}=0.2$ ($\alpha=0.8$) a two
step DMRG calculation\cite{Moukouri06} shows a much wider non magnetic
region centered around the classical transition point $J_2/J_{1y}=0.1$.
Interestingly, the spin gap at this maximally frustrated point is almost
identical to that of an isolated Haldane chain suggesting that even for
relatively strong interchain couplings one dimensional Haldane chain physics
is still important. This is certainly missed by the SBMF calculation which
treats the spin as a continuous variable and does not distinguish between
integer and half-integer spins crucial for the existence of the Haldane spin
gap. Moreover, it has been established within quantum Monte Carlo
calculations\cite{Kim+00,Matsumoto+01,Alet+02} that for $J_2=0$ the
transition between the N\'eel ordered phase and the gapped non-magnetic
phase is located at $J_{1x}/J_{1y}=0.044$, again indicating that the
paramagnetic region close to the Haldane chain limit is considerably wider
than suggested by both the SBMFT and the coupled cluster results\cite%
{Bishop+08}. Since for $J_2=0$ the system is not frustrated quantum Monte
Carlo can be considered exact in this regime.

We have also calculated the transition out of the $(0,\pi)$ state in linear 
spin-wave theory, for comparison to the SBMFT calculation and to preliminary
neutron scattering results on the pnictides that found $(0,\pi)$ order but
with a small moment. This method computes the reduction of the classical
antiferromagnetic moment due to zero-point excitations of spin waves (which
captures the $1/S$ correction to the classical moment in a large-$S$
expansion). The transition line out of the ordered phase is estimated as the
point where the correction is as large as the original moment.

We assume three antiferromagnetic couplings: $J_{1x}$, $J_{1y}$, and $J_{2}$ 
with $J_{1y}>J_{1x}$ and that we are in a $(0,\pi )$ ordered
phase, which requires (as is evident from the formula below) that $%
J_{1x}<2J_{2}$. The dispersion relation for spin-wave excitations was
previously obtained for this anisotropic $J_{1}$-$J_{2}$ model in Ref. %
\onlinecite{Kruger+08}. The integral for the correction to the classical
moment around the $(0,\pi )$ case is, in units of the Bohr magneton and with
lattice spacing $a=1$,

\begin{equation}
\Delta m = \int_{[0,2\pi]^2} \frac{\textrm{d}^2 \mathbf{k}}{(2\pi )^2}
\left( \frac{1}{\sqrt{1-\cos ^{2}(k_{y})/f(k_{x})^{2}}}-1\right)\quad
\end{equation}
with
\begin{equation}
f(k_{x})=\frac{2J_{2}+J_{1y}-J_{1x}[1-\cos (k_{x})]}{2J_{2}\cos
(k_{x})+J_{1y}}.
\end{equation}

The transition is found numerically to lie very close to the classical
transition line $J_{1x}=2J_{2}$: the normalized difference $%
(J_{2}^{c}-J_{1x}/2)/J_{1y}$, where $J_{2}^{c}$ is the critical coupling
where the correction is equal to the original moment, is always less than
2\% for $0<J_{1x}<0.99J_{1y}.$ Significant reduction of the ordered moment
also occurs only near the classical transition line.
The primary difference from the SBMFT calculation is that the spin-wave
calculation always predicts a paramagnetic phase between stripe and Ne\'{e}l
order. The width of this paramagnetic phase is much smaller in either
approach than in the DMRG calculation of the following section, because
these analytical approaches do not capture the strong quantum fluctuations
that favor the Haldane phase.

\section{DMRG}

\label{sec.DMRG}

The above SBMFT has clearly shown an interesting narrow boundary region
between the N\'eel and stripe ordered phases in the phase diagram of Fig.~%
\ref{Fig:SBMFT}, where the quantum fluctuations are expected to become very
important. Most interestingly, it suggests an increasing tendency towards a
fluctuation induced first order transition on approaching the isotropic
point $J_{1x}=J_{1y}$. However, the comparison with previous numerical
results\cite{Moukouri06,Kim+00,Matsumoto+01,Alet+02} indicates that the
SBMFT tends to overestimate the stability of the magnetically ordered
phases, surely close to the Haldane chain limit but presumably also for
larger values of $J_{1x}$ and $J_2$. In the following we shall refine the
boundary region by using DMRG method.

In the following DMRG calculation, we will set $J_{1y}=J_{1}$ as energy
unit, and a periodic boundary condition (PBC) is used and in each DMRG block
up to $m=3200$ states are to be kept with the truncation error in the order
of or less than $10^{-5}$.

\subsection{Isotropic case with $J_{1x}= J_{1y}= J_1$}

\begin{figure}[h]
\centerline{
    \includegraphics[height=2.0in,width=3.6in]{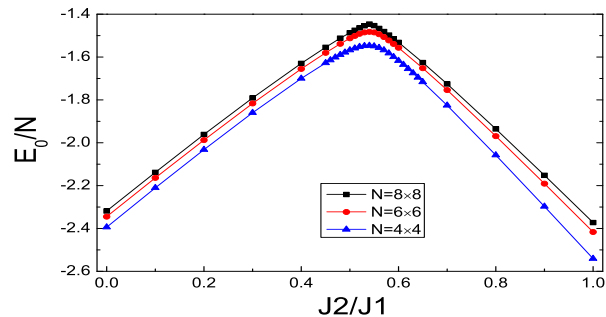}
    }
\caption{(color online) The ground-state energy per site at
different system size $N=4\times 4$ (blue triangle), $6\times 6$
(red circle) and $8\times 8$ (black square), for the isotropic case
with $J_{1x}=J_{1y}=J_{1}$. } \label{Fig:Eg_J1x1}
\end{figure}

Let us first consider the isotropic case $J_{1x}=J_{1y}=J_{1}$ where the
SBMFT suggests the strongest tendency towards a first order transition
although the transition point $J_{2}/J_{1}\approx 0.64$ obtained from a
comparison of the energy minima seems to be suspiciously high compared to
the classical transition point $J_{2}/J_{1}=0.5$. Fig. \ref{Fig:Eg_J1x1}
shows the ground state energy per site calculated by DMRG with the sample
size varying from $N=4\times 4$ ($16$ sites) up to $N=8\times 8$ ($64$
sites). The ground state energy reaches the maximum with $J_{2}$ between $%
0.54J_{1}$ and $0.58J_{1}$, which becomes sharper with the increase of
sample size, indicating a region with possible phase transitions below the
first order transition point obtained in SBMFT but still notable above the
classical transition.

\begin{figure}[tbp]
\centerline{
    \includegraphics[height=3.0in,width=1.0\columnwidth]{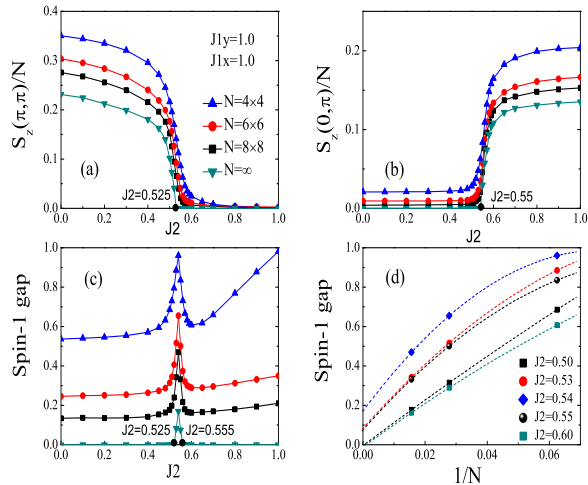}
    }
\caption{(color online) The evolution of the peak values of the structure
factors, $S_{z}(\protect\pi ,\protect\pi )/N$ (a), $S_{z}(0,\protect\pi )/N$
(b), and of the spin-1 gap (c), at three different size $N=4\times 4$, $%
6\times 6$ and $8\times 8$, as well as their thermodynamic limit
extrapolations in the isotropic case $J_{1x}=J_{1y}=J_{1}=1$. The
finite-size scaling for the spin-1 gap is shown in (d). }
\label{Fig:J1x1_SFac}
\end{figure}

We examine such a region by calculating the magnetic structure factor $S_{z}(%
\mathbf{q})$, defined by

\begin{equation*}
S_{z}(\mathbf{q})={\frac{1}{N}}{\sum_{i,j}}e^{-i\mathbf{q}(\mathbf{r}_{i}-%
\mathbf{r}_{j})}\langle S_{i}^{z}S_{j}^{z}\rangle .
\end{equation*}%
As expected, we find that $S_{z}(\mathbf{q})$ shows a dramatic change from
peaking at ${\bm Q}=(\pi ,\pi )$ (the N\'{e}el order) to ${\bm Q}=(0,\pi )$
(the stripe order) with increasing $J_{2}/J_{1}$. Fig. \ref{Fig:J1x1_SFac}
(a) and (b) illustrate $S_{z}(\mathbf{q})/N$ vs. $J_{2}/J_{1}$ for system
sizes $N=4\times 4$, $6\times 6$ and $8\times 8$, as well as the
thermodynamic limit values obtained by a finite size scaling. Here we have
used the quadratic function $f(x)=A+Bx+Cx^{2}$ with $x=\frac{1}{N}$ to
perform the finite-size scaling.

The N\'{e}el order transition point is found at $J_{2}/J_{1}\approx 0.525$
in Fig. \ref{Fig:J1x1_SFac}(a), which is clearly distinct from the stripe
order tranistion point at $J_{2}/J_{1}\approx 0.55$ in Fig. \ref%
{Fig:J1x1_SFac}(b) indicating that the two magnetically ordered phases are
separated by an intermediate non-magnetic region.

To independently verify the above results, we also calculate the spin-1 gap $%
\Delta E(S=1)\equiv E_{1}(S=1)-E_{0}$ presented in Fig. \ref{Fig:J1x1_SFac}%
(c) and (d). Fig. \ref{Fig:J1x1_SFac}(d) illustrates the finite-size scaling
for the spin gap $\Delta E(S=1)$ at different values of $J_{2}$ using a
scaling function $f(x)$. In Fig. \ref{Fig:J1x1_SFac}(c), the evolution of
the spin-1 gap as a function of $J_{2}$ is given. In the thermodynamic
limit, the transition points determined by the spin-1 gap are at $%
J_{2}\approx 0.525$ and $J_{2}\approx 0.555$, respectively, which are very
close to the previous results determined by the structure factor. Therefore,
for the present $S=1$ $J_{1}$-$J_{2}$ model in the isotropic limit, our
numerical approach has established an intermediate spin disordered region
with a finite spin gap which separates the two ordered magnetic phases.

\subsection{Anisotropic case with $J_{1x}<J_{1y}=J_{1}$}

\begin{figure}[tbp]
\centerline{
    \includegraphics[height=3.0in,width=1.0\columnwidth]{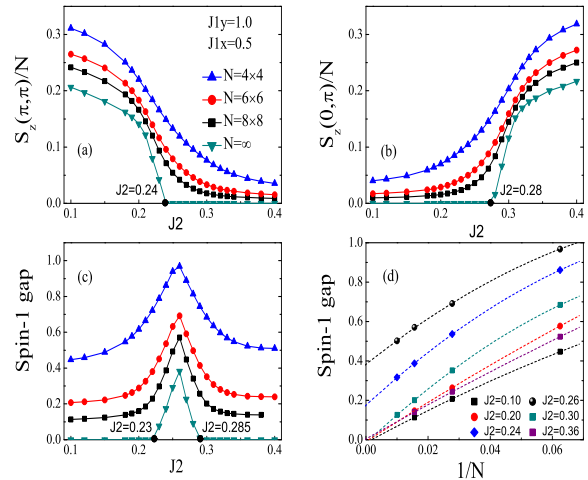}
    }
\caption{(color online) The peak values, (a) $S_{z}(\protect\pi ,\protect\pi %
)/N,$ (b) $S_{z}(0,\protect\pi )/N$, and (c) the spin-1 gap, versus $%
J_{2}/J_{1y}$ at $J_{1x}/J_{1y}=0.5$ for different sizes$.$ The finite-size
scaling for the spin-1 gap is shown in (d). }
\label{Fig:SFacJ1x05}
\end{figure}

Now we consider how the spin disordered phase evolves with the increase of
anisotropy at $J_{1x}<J_{1y}=J_{1}$. First we consider the case at $%
J_{1x}=0.5J_{1}$, and the results are presented in Fig. \ref{Fig:SFacJ1x05}.
By using the same finite size scaling procedure, we find that the spin
disordered phase is bound by a lower transition point $J_{2}/J_{1}\approx
0.24$ and an upper transition point $J_{2}/J_{1}\approx 0.28$ based on the
structure factor calculation. Again the spin gap calculation gives rise to a
consistent spin disordered regime between $J_{2}/J_{1}\approx 0.23$ and $%
J_{2}/J_{1}\approx 0.285$ as shown in Fig. \ref{Fig:SFacJ1x05} (c).

In the same regime, with a fixed $J_{2}/J_{1}=0.25$, we have further studied
the phase boundaries by varying $J_{1x}/J_{1}$. As shown in Fig. \ref%
{Fig:SFacJ2q025}, the lower transition point is obtained at $%
J_{1x}/J_{1}=0.45$ and the upper transition point at $J_{1x}/J_{1}\approx
0.52$, while the finite size scaling for the spin-1 gap results in a similar
region between $J_{1x}/J_{1}=0.44-0.54$.

\begin{figure}[tbp]
\centerline{
    \includegraphics[height=3.0in,width=1.0\columnwidth]{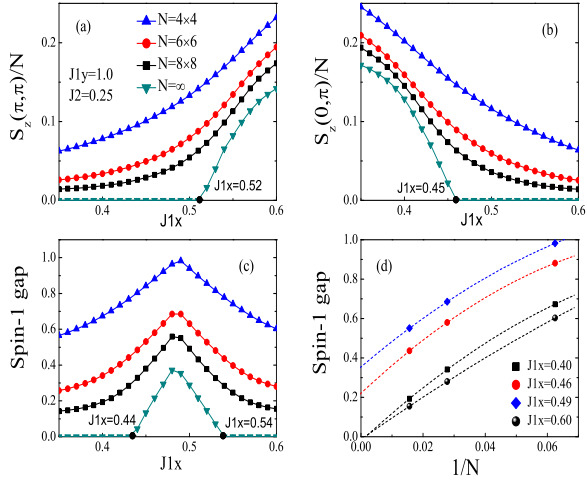}
    }
\caption{(color online) (a) $S_{z}(\protect\pi ,\protect\pi )/N$, (b) $%
S_{z}(0,\protect\pi )/N$, and (c) the spin-1 gap, versus $J_{1x}/J_{1y}$ at $%
J_{2}/J_{1y}=0.25$ for different sample sizes. In (d), the finite-size
scaling for the spin-1 gap is given. }
\label{Fig:SFacJ2q025}
\end{figure}
\begin{figure}[tbp]
\centerline{
    \includegraphics[height=3.0in,width=3.6in]{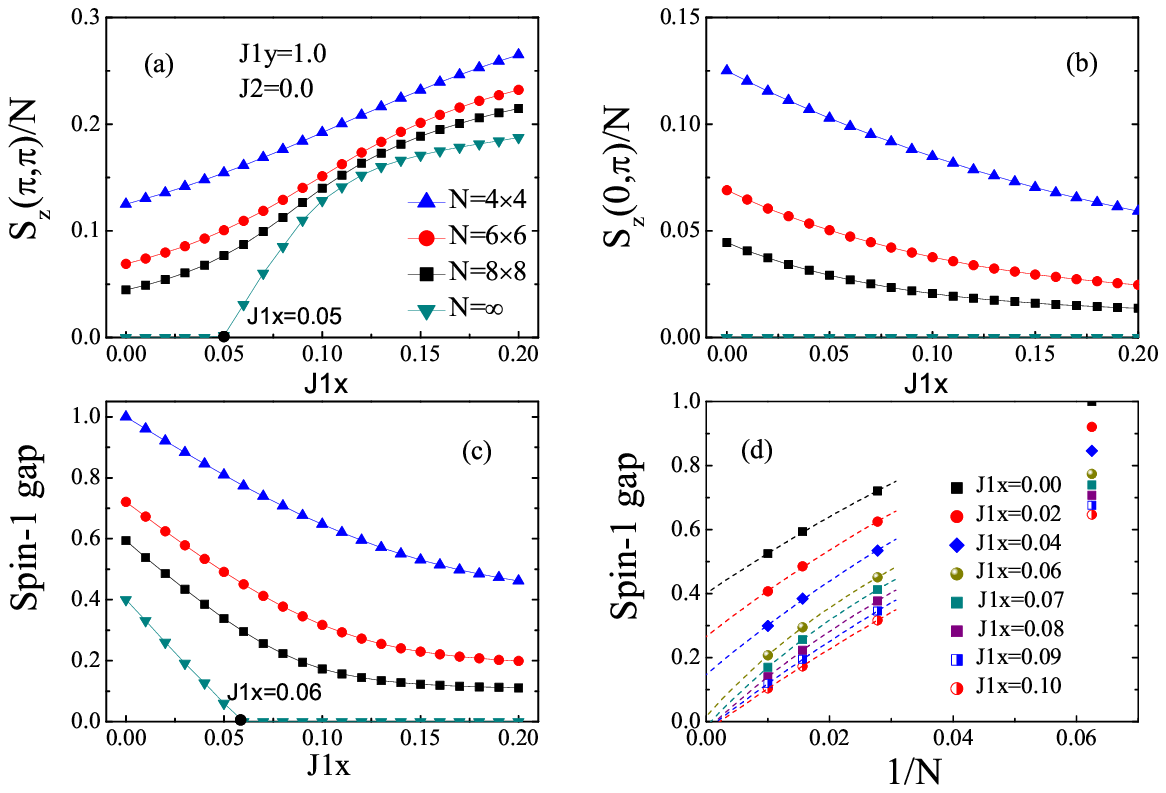}
    }
\caption{(color online) (a) $S_{z}(\protect\pi ,\protect\pi )/N$, (b) $%
S_{z}(0,\protect\pi )/N$, and (c) the spin-1 gap, versus $J_{1x}/J_{1y}$ at $%
J_{2}=0$ at different sizes. In (d), the finite-size scaling for the spin-1
gap is given. }
\label{Fig:SFacJ2q00}
\end{figure}
\begin{figure}[tbp]
\centerline{
    \includegraphics[height=3.0in,width=1.0\columnwidth]{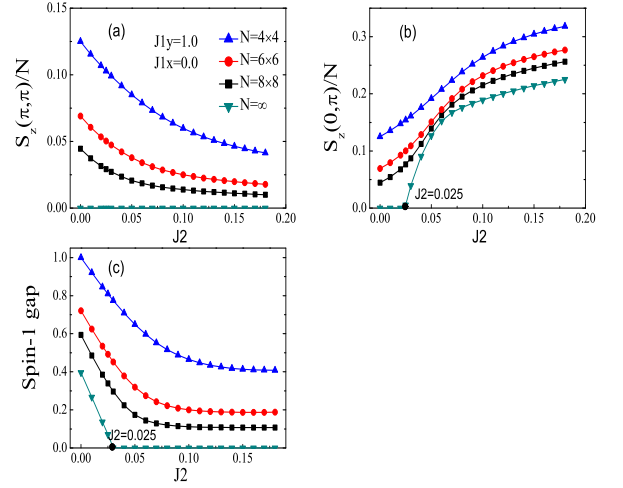}
    }
\caption{(color online) At $J_{1x}=0$, the structure factors $S_{z}(\protect%
\pi ,\protect\pi )/N$ (a) and $S_{z}(0,\protect\pi )/N$ (b) as well as the
spin-1 gap (c) versus $J_{2}/J_{1y}$ are shown at different sizes, including
their thermodynamic extrapolations. }
\label{Fig:SFacJ1x00}
\end{figure}
\begin{figure}[tbp]
\centerline{
    \includegraphics[height=3.0in,width=1.0\columnwidth]{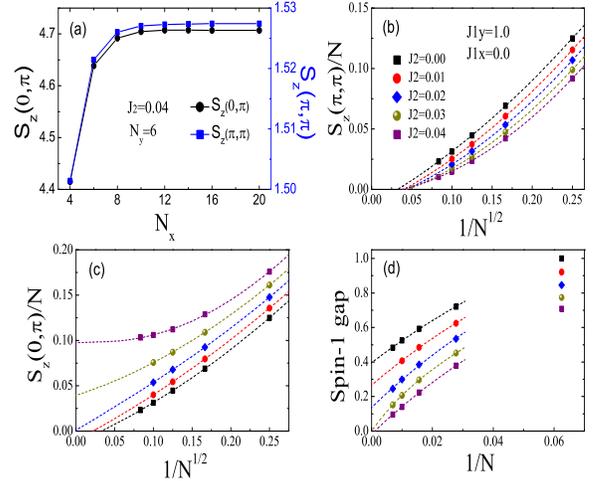}
    }
\caption{(color online) (a) The size dependence of the structure factor [for
a given $N_{y}=$ $6$ (6-leg systems)] at $J_{1x}=0$ and small $J_{2}/J_{1y}$
which quickly saturates at $N_{x}>N_{y}$, indicating that $x=1/N_{y}$ in the
scaling function $f(x)$ for the structure factor is more appropriate in this
extreme limit. The corresponding finite-size scaling of the structure
factors for square samples are illustrated in (b) and (c), respectively. For
comparsion, the spin-1 gap at different $J_{2}$ is still well scaled by $%
x=1/N$ in the scaing function as shown in (d). }
\label{Fig:J1x00_Scaling}
\end{figure}

In the extreme case at $J_{1x}=0$ and $J_{2}=0$ with $J_{1y}=J_{1}$, the
system simply reduces to an array of decoupled $S=1$ spin chains with a
finite Haldane gap. Fig. \ref{Fig:SFacJ2q00} shows how the ground state
continuously evolves from that of the well-known decoupled spin chains to
the anisotropic 2D case by turning on $J_{1x}/J_{1}$, which remains
disordered until the N\'{e}el order sets in at $J_{1x}/J_{1}\approx 0.05$
with $S_{z}(\pi ,\pi )/N$ becoming finite in the thermodynamic limit. Indeed
the corresponding spin-1 gap continuously decreases with the turning on of a
finite $J_{1x}$, but only vanishes around $J_{1x}/J_{1}\approx 0.06$ as
shown in Fig. \ref{Fig:SFacJ2q00} (c) and (d).

Now we turn on $J_{2}$. At $J_{1x}=0,$ we find that while the calculated
structure factor at $(\pi ,\pi )$ continuously reduces as the sample size
increases from $N=4\times 4$, $6\times 6$, to $8\times 8$, and is
extrapolated to zero by finite size scaling, a finite stripe order $%
S_{z}(0,\pi )/N$ will emerge at\ $J_{2}/J_{1y}=0.025$ in the thermodynamic
limit, which is further supported by vanishing spin-1 excitation gap at the
same point as shown in Fig. \ref{Fig:SFacJ1x00}. It is noted that the best
finite scaling for the structure factor here is obtained with using a $f(x)$
with $x=1/\sqrt{N}=1/N_{y},$ for a square lattice $N_{x}=N_{y},$ instead of $%
x=1/N$ used previously. The justification of such a finite-size scaling for
the spin structure factors at $J_{1x}=0$ and small $J_{2}/J_{1y}$ is given
in Fig. \ref{Fig:J1x00_Scaling} and its caption.

\subsection{Phase diagram}

\begin{figure}[tbp]
\centerline{
    \includegraphics[height=2.4in,width=3.6in]{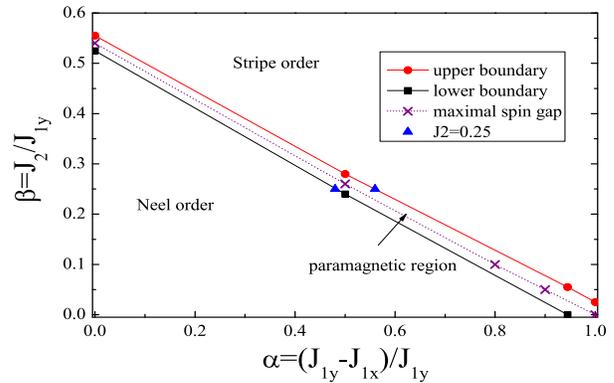}
    }
\caption{(color online) Phase diagram for the anisotropic $J_{1}$-$J_{2}$
model determined by DMRG. The N\'{e}el order and stripe order phases are
separated by a paramagnetic regime with the boundaries denoted by the red
line with full circles for the upper and the blank line with full squares
for the lower transition points, respectively. In the middle of the
paramagnetic region lies in a dotted line with crosses which marks the
maximal spin-1 gap $\Delta E_{\mathrm{max}}$. Note that the symbols of
circle, square, and triangular denote the phase transition points determined
by DMRG in the previous figures. }
\label{Fig:PhaseDiagram}
\end{figure}

The resulting phase diagram obtained by the DMRG calculations with careful
finite size scalings of the magnetic structure factor as well as the spin
gap is shown in Fig. \ref{Fig:PhaseDiagram} as a function of the anisotropy $%
\alpha =(J_{1y}-J_{1x})/J_{1y}$ between the NN couplings and the relative
strength $\beta =J_{2}/J_{1y}$ of the NNN superexchange.

In contrast to the SBMFT which in the regime of small anisotropy $\alpha $
suggests a first order transition between the N\'{e}el and stripe ordered
phases (see Fig.~\ref{Fig:SBMFT}), the DMRG results clearly indicate a
paramagnetic strip separating the two magnetically ordered phases for all
values of the anisotropy $\alpha $ including the isotropic point $%
J_{1x}=J_{1y}$ ($\alpha =0$). With increasing $\alpha $ the width of this
region is found to slightly increase. Close to the Haldane chain limit the
SBMFT predicts an intermediate non magnetic phase, though the width of this
region is found to be much larger in the DMRG calculation in quantitative
agreement with a previous two step DMRG analysis at fixed $\alpha =0.8$
(Ref.~\onlinecite{Moukouri06}).

For $J_{2}=0$ we can compare our results to recent QMC simulations\cite%
{Kim+00,Matsumoto+01,Alet+02}, which in this regime can be considered exact
since the system is not frustrated and QMC is therefore free of any sign
problem. Within DMRG we find the transition between the gapped Haldane phase
and the N\'{e}el ordered phase to occur at $J_{1x}/J_{1y}\approx 0.05$ (see
Fig.~\ref{Fig:PhaseDiagram}) in good agreement with QMC transition point $%
J_{1x}/J_{1y}=0.044$. Moreover, in the Haldane chain limit $J_{1x}=J_{2}=0$
we obtain a spin gap $\Delta \approx 0.4$ (see Fig.~\ref{Fig:SFacJ2q00}(d))
very close to the exact value $\Delta _{H}=0.41$ for the Haldane spin chain%
\cite{Sorenson+93}, again demonstrating that the finite size scaling is well
converged.

Very interestingly, starting from the Haldane chain limit we find the
maximum spin gap in the paramagnetic phase to decrease only very slowly with
increasing couplings $J_{1x}$ and $J_{2}$. Up to $J_{1x}/J_{1y}=0.5,$ the
gap is almost identical to the Haldane spin gap (see Fig.~\ref{Fig:SFacJ1x05}%
(d)) indicating that the Haldane spin chain is very robust against the
simultaneous increase of the mutually frustrating couplings $J_{1x}$ and $%
J_{2}$. Remarkably, even the paramagnetic phase of the isotropic $J_{1}$-$%
J_{2}$ model with a maximum spin gap still being about half of the Haldane
gap is continuously connected to the Haldane spin-chain limit.

At the isotropic point $J_{1x}=J_{1y}$ ($\alpha=0$) we find an intermediate
paramagnetic phase for $0.525\lesssim J_2/J_1\lesssim 0.555$. This region is
considerably smaller than in the $S=\frac 12$ case, where a paramagnetic
regime $0.4\lesssim J_2/J_1\lesssim 0.6$ presumably with columnar valence
bond order has been established by various numerical methods\cite%
{Dagotto+89,Schulz+92,Schulz+96,Sindzingre04,
Capriotti+00,Capriotti+01,Gerlfand+89,Gerlfand90,Singh+99,Sushkov+01,Sushkov+02}%
. Interestingly, the stabilization of the N\'eel phase above the classical
transition point is in agreement with the SBMFT.

Despite the apparent failure of the semiclassical SBMFT in dealing with the
strong quantum fluctuations in the boundary region between the two
magnetically ordered phases it is interesting to note that the paramagnetic
region covers the phase boundaries obtained by SBMFT for the most of $%
J_{1x}<J_{1y}=J_{1}$ region except for the part close to the isotropic limit
and that the points where the spin gap is maximum (indicated by crosses and
dotted in Fig.~\ref{Fig:PhaseDiagram}) are almost on top of both the first
order transition line as well as of the two narrow second order lines
obtained within SBMFT.

\section{Discussion}

\label{sec.discussion}

In summary, we have studied the frustrated spin-1 Heisenberg $J_{1x}$-$%
J_{1y} $-$J_2$ model on a square lattice numerically using the DMRG method
and analytically employing the Schwinger-Boson mean-field theory (SBMFT).
Interestingly, this model contains both the isotropic $J_1$-$J_2$ model as
well as decoupled Haldane spin chains as limiting cases. Moreover, it has
attracted a lot of attention recently since it has been motivated as an
effective model to describe the $(0,\pi)$-magnetism and the low-energy
spin-wave excitations of the celebrated novel iron pnictide superconductors.
Furthermore it has been suggested that the drastic reduction of magnetic
moments to a value of 0.4 $\mu_B$ is caused by strong quantum fluctuations
in the vicinity of a continuous phase transition. However, to the present
day the phase diagram has not been determined yet.

Within both the SBMFT and the DMRG we find that the N\'{e}el phase is
stabilized considerably by quantum fluctuations above the classically stable
region up to a relatively strong anisotropy between the NN couplings.
However, in all other regards the phase diagrams obtained by the two methods
clearly disagree, indicating the importance of the strong quantum
fluctuations.

The SBMFT suggest a fluctuation induced first order transition between the
N\'eel and the stripe antiferromagnets terminating at a tricritical point
and splitting into two second order lines separated by an intermediate
paramagnetic region only for large anisotropies. Although the existence of a
tricritical point and a hardly seizable paramagnetic strip terminating at
the decoupled Haldane spin chain point ($J_{1x}=J_2=0$) are consistent with a
recent coupled cluster calculation\cite{Bishop+08}, the SBMFT definitely falls short on
approaching the Haldane chain limit. This becomes clear by a comparison with
quantum Monte Carlo simulations\cite{Kim+00,Matsumoto+01,Alet+02}
showing a much larger region of stability of
the Haldane chain phase against interchain couplings $J_{1x}$. Also the
previous two-step DMRG calculation\cite{Moukouri06} at $J_{1x}/J_{1y}=0.2$ indicates a
paramagnetic region being an order of magnitude wider than found within the
SBMFT.

On the contrary, the phase diagram obtained within DMRG by a careful finite
size scaling of the magnetic structure factor and the spin gap is consistent
with both the QMC and the previous DMRG results. Furthermore the spin gap in
the decoupled chain limit agrees well with the exact value for the isolated
Haldane chain. The width of the paramagnetic region slightly decreases on
approaching the isotropic point but remains finite over the entire parameter
range. This has the remarkable implication that the paramagnetic phase of
the frustrated two dimensional spin model, including the isotropic $J_{1}$-$%
J_{2}$ limit, is continuously connected to the Haldane spin-chain phase. In
other words, the paramagnetic phase can be viewed as a continuation of the
Haldane spin chain phase,  with the caveat that the topological string order
has to disappear at any finite interchain coupling\cite{Todo+01,Anfuso+07}.

The reason for the failure of the semiclassical SBMFT in dealing correctly
with the strong quantum fluctuations in the boundary region between the two
classical orders is easy to understand. The SBMFT deals with the spin
as continuous variable while it is {\em blind for the Berry phase effects} that
distinguish between half-integer and integer spin values crucial for the existence of the
Haldane gap as well the valence bond crystals. This physics is definitely missed
by the semiclassical treatment. Our DMRG results demonstrates that the Haldane chain phase is very
robust against the simultaneous increase of the couplings $J_{1x}$ and $%
J_{2} $ along the strongly frustrated boundary region, indicated by the minute reduction 
of the spin gap compared to the isolated Haldane chain limit.

What does the study of this spin system teach us regarding the
superconductivity in the iron pnictides? One could speculate that
the basic physics is similar as in the cuprates. Although the spins
are larger in the pnictides, the geometrical frustration renders the
spin system to be on the verge of undergoing a quantum phase
transition in a quantum disordered state. This quantum spin physics
then sets the conditions for the emergence of superconductivity in
the doped systems. But in this regard the size of the microscopic
spin does matter more than one intuitively anticipates. For $S=1/2$
the well established fact that the Berry phases conspire to turn the
quantum disordered states of the spin-only systems into valence bond
solids gives a rationale to take Anderson's resonating valence bond
(RVB) idea for the origin of  high Tc superconductivity quite
seriously. The valence bonds are protected by the spin gap and the
effect of doping could well be to just turn the valence bond solids
into translational quantum liquids that transport  two units of
electrical charge. Focussing on the Berry phases in the pnictides
one has only the options of an `intermediate' crystal field $S=1$
state\cite{Kruger+08} or the high spin $S=2$ for the microscopic
spin. In the former case the ground state has a twofold degeneracy
and the `building blocks' are no longer  pair singlets but instead
the `chain singlets'. Thinking along the RVB lines, what to expect
when such a system is doped? The dopants will increase the quantum
fluctuations but chains are not pairs. One anticipates that doping
might drive the system into the non-magnetic `Haldane chain phase'
breaking the two-fold rotational symmetry of the lattice. One
notices that something of the kind is found in the phase diagram of
the pnictides: the structural transition to the orthorombic phase
persists to much higher dopings than the stripe antiferromagnetism\cite{Zhao+08b}. 
At first view this seems
rather detached from the RVB idea. But now one has to realize that
the charge carriers are themselves spinfull, carrying by default a
half-integer spin. Taking for instance the intermediate $S=1$
background, the holes would carry $S=1/2$, and t-J models
corresponding with a mix of $S=1$ and $S=1/2$ states have been
studied in the past\cite{Zaanen+92}. One anticipates that in an
incompressible `chain like' $S=1$ background the $S=1/2$ carriers
might again be `glued by Berry forces' into valence bond pairs,
re-establishing a connection with the RVB mechanism of the cuprates.
A similar idea has been previously proposed from a different
approach\cite{Baskaran08}.

In conclusion, it is quite questionable that any of these considerations have a bearing on pnictide superconductivity but they
do make the case that there is still much interesting physics to be explored dealing with systems characterized by a larger
microscopic spin.

\textbf{Acknowledgment:} We are grateful for stimulating discussion with M.
Q. Weng. This work is partially supported by NSFC and the National Program
for Basic Research of MOST, China, the DARPA OLE program, the DOE grant
DE-FG02-06ER46305,  the NSF grant DMR-0605696 and the Stichting voor
Fundamenteel Onderzoek  der Materie (FOM).

\end{document}